\documentclass{ws-ijmpa}

\begin{document}

%%%%%%%%%%%%%%%%%%%%%%%%%%% To switch off trimmarks %%%%%%%%%%%%%%%%%%%
%

\def\nocropmarks{\vskip5pt\phantom{cropmarks}}

\let\trimmarks\nocropmarks      %%% Pls. remove the comment sign (%) to switch off the trimmarks

%
%%%%%%%%%%%%%%%%%%%%%%%%%%%%%%%%%%%%%%%%%%%%%%%%%%%%%%

\markboth{X.-Q. Li, X.-B. Zhang \& B.-Q. Ma} {Strange Sea
Asymmetry in Nucleons}

%%%%%%%%%%%%%%%% Publisher's Area please ignore %%%%%%%%%%%%%%%
%
\catchline{}{}{}
%
%%%%%%%%%%%%%%%%%%%%%%%%%%%%%%%%%%%%%%%%%%%%%%%%%

\setcounter{page}{1}

\title{%Asymmetry of Strange Sea in Nucleons
STRANGE SEA ASYMMETRY IN NUCLEONS\footnote{ Talk presented at the
Third Circum-Pan-Pacific Symposium on ``High Energy Spin Physics",
Oct.~8-13, 2001, Beijing, China. This work is partially supported
by the National Natural Science Foundation of China. } }

\author{\footnotesize XUE-QIAN LI$^{1}$, XIAO-BING ZHANG$^{1}$ and BO-QIANG
MA$^{2}$}
%\footnote{
%Typeset names in
%10 pt roman, uppercase. Use the footnote to indicate the
%present or permanent address of the author.}}

\address{$^{1}$Department of Physics, Nankai University, Tianjin 300071,
China\\
\vspace{0.6mm} $^{2}$Department of Physics, Peking University,
Beijing 100871, China}
% \footnote{State completely without abbreviations, the
%affiliation and mailing address, including country. Typeset in 8
%pt italic.} }

%\author{SECOND AUTHOR}

%\address{Group, Laboratory, Address\\
%City, State ZIP/Zone, Country
%}

\maketitle

%\pub{Received (received date)}{Revised (revised date)}

\begin{abstract}
We evaluate the medium effects in nucleon which can induce an
asymmetry of the strange sea. The short-distance effects
determined by the weak interaction can give rise to $\delta
m\equiv \Delta m_s-\Delta m_{\bar s}$ where $\Delta m_{s(\bar s)}$
is the medium-induced mass of strange quark by a few KeV at most,
but the long-distance effects by strong interaction could be
sizable.
\end{abstract}

%\section{General Appearance}    %) A SECTION HEADING

\section*{~}
\vspace{-7mm}

%\noindent (1)
%The existence of the Dirac sea is always an interesting topic
%which all theorist and experimentalists of high energy physics are
%intensively pursuing, and the strange content of the nucleon sea
%is of particular interest for attention.
The strange content of the nucleon is under particular attention
by the high energy physics society recently. Ji and Tang\cite{Ji}
suggested that if a small locality of strange sea in nucleon is
confirmed, some phenomenological consequences can be resulted in.
The CCFR data\cite{CCFR} indicate that $s(x)/\bar s(x)\sim
(1-x)^{-0.46\pm 0.87}$. Assuming an asymmetry between $s$ and
$\bar s$, Ji and Tang analyzed the CCFR data and concluded that
$m_s=260\pm 70$ MeV and $m_{\bar s}=220\pm 70$ MeV\cite{Ji}. So if
only considering the central values, $\delta m\equiv m_s-m_{\bar
s}\sim 40$ MeV. In the framework of the Standard Model
$SU(3)_c\otimes SU(2)_L\otimes U(1)_Y$, we would like to look for
some possible mechanisms which can induce the asymmetry.

The self-energy of strange quark and antiquark $\Sigma_{s(\bar
s)}=\Delta m_{s(\bar s)}$ occurs via loops where various
interactions contribute to $\Sigma_{s(\bar s)}$ through the
effective vertices. Obviously, the QCD interaction cannot
distinguish between $s$ and $\bar s$, neither the weak interaction
alone in fact. Practical calculation of the self-energy also shows
that $\Delta m_s=\Delta m_{\bar s}$. In fact, because of the CPT
theorem, $s$ and $\bar s$ must be of exactly the same mass.

If we evaluate the self-energy $\Delta m_s$ and $\Delta m_{\bar
s}$ in vacuum, the CPT theorem demands $\Delta m_s\equiv \Delta
m_{\bar s}$. However, when we evaluate them in an asymmetric
environment of nucleons, an asymmetry $\Delta^{M} m_s\neq
\Delta^{M} m_{\bar s}$ where the superscript $M$ denotes the
medium effects, can be expected. In other words, we suggest that
the asymmetry of the $u$ and $d$ quark composition in nucleons
leads to an asymmetry of the strange sea.

There exist both short-distance and long-distance medium effects.
The short-distance effects occur at quark-gauge boson level,
namely a self-energy loop including a quark-fermion line and a
W-boson line or a tadpole loop. The contributions of $u$ and
$d$-types of quark-antiquark to the asymmetry realize through the
Kabayashi-Maskawa-Cabibbo mixing.

The calculations at the parton-$W(Z)$ level is trustworthy,
because it is carried out in the standard framework which has been
proved to be correct. One does not suspect its validity and trusts
that this mechanism can cause an asymmetry of the strange sea in
nucleons. However, later we will show that it can only result in a
$\delta m$ of order of a few KeV, much below what we need for
phenomenology.

Accepting the value of $\delta m$ achieved by fitting data as
about 40 MeV, one has to look for  other mechanisms which can
bring up larger $\delta m$. Obviously the smallness of $\delta m$
is due to the heavy $W$ or $Z$ bosons  in the propagators. and
they are responsible for the weak interaction. We would ask if the
strong interaction can get involved, if yes, it definitely
enhances the $\delta m$ by orders. However, the parton-gluon
interaction cannot lead to the asymmetry, because gluon is
flavor-blind. Thus the perturbative QCD where gluons are exchanged
between partons does not apply in this case. A natural extension
would be that the long-distance interaction may result in a larger
asymmetry. It is generally believed that the long-distance effects
exist at the quark level, but the realm is fully governed by the
non-perturbative QCD, so the question is how to evaluate the
long-distance effects.

In fact, Brodsky and one of us proposed a meson-baryon resonance
mechanism and they suggested that the sea quark-antiquark
asymmetries are generated by a light-cone model of energetically
favored meson-baryon fluctuations\cite{Brodsky}.

We try to re-evaluate the asymmetry from another angle, namely, we
consider the interaction of quark(parton)-meson. Here there is a
principal problem that the parton picture was introduced for high
energy processes where partons are treated massless compared to
the involved energy scale. That is an self-consistent picture
where the chiral symmetry is respected. Can the picture enclose
the quark-meson interaction is still a puzzle. But as the
phenomenology suggests, the long-distance strong interaction
should apply in this case, there can be possibility to treat the
quark-meson interaction as for the constituent-quark-meson
interaction, even though at this energy scale (the invariant
masses of the mesons) the chiral symmetry is broken. There is
another reason to believe the picture that the pseudoscalar mesons
$\pi$, $K$ etc. are composite of SU(3) quarks and antiquarks, but
also are the Goldstine bosons, so they must satisfy the
Bethe-Salpeter equation and the Dyson-Schwinger equation
simultaneously. The picture may become self-consistent when the
non-perturbative  QCD effects can be properly regarded. At this
stage we just postulate that we can apply the chiral lagrangian to
treat the quark-meson interaction where the sea quark(antiquark)
and valence quarks are all included.

Many authors employed this scenario to estimate various flavor
asymmetries and spin contents\cite{eich,cheng,szcz}, where the sea
quarks(antiquarks) make substantial contributions. However, in
Ref.~4,
%\cite{eich},
the constituent quark mass of 340 MeV was employed, whereas, in
Ref.~5,
%\cite{cheng},
the current quark mass relation $m_s/\hat
m=25$ is used where $\hat m$ is the mass of the light quarks ($u$
and $d$). This discrepancy still comes from lack of solid
knowledge on the non-perturbative QCD. In our work, we vary the
quark masses and see how the numerical values change. Our results
indicate that the difference for various quark masses is not too
remarkable.

For the valence and sea quark picture, one has to use the quark
distribution function which has obvious statistical meaning. Here
instead of the commonly used distribution function, we adopt the
distribution with finite medium temperature and density. The
temperature involved in the distribution is only a parameter which
characterizes the inner motion state of the quarks (valence and
sea) and has the order of $\Lambda_{QCD}$. In practice, we let the
temperature vary within a reasonable range $100 \to 300$ MeV. The
advantage of using the finite temperature field theory is obvious.
First, the theory is well-established and then the calculations
are simple and straightforward.

%\\
%\noindent (2)
We are going to employ the familiar formulation of the Quantum
Field Theory at finite temperature and density. As well-known, the
thermal propagator of quarks can be written as %\cite{ftft}
\begin{equation}
\label{pro} iS_q(k)=\frac{i(\rlap /k+m_q)}{k^2-m_q^2}-2\pi(\rlap
/k+m_q)\delta(k^2-m_q^2)f_F(k\cdot u),
\end{equation}
where $u_\mu$ is the four-vector for the medium and $f_F$ denotes
the Fermi-Dirac distribution function
\begin{equation}\label{ff}
f_F(x)={\theta(x)\over e^{\beta(x-\mu)}+1}+{\theta(-x)\over
e^{-\beta (x-\mu)} +1},
\end{equation}
and $\beta=1/kT$, $\mu$ is the chemical potential. We notice that
the first term of Eq.~(\ref{pro}) is just the quark propagator in
the vacuum. Its contribution to $\Sigma_1$ is of no importance to
us because this is related to the wave-function renormalization of
the quark in the vacuum. We focus on the medium effect, which
comes from the second term of Eq.~(\ref{pro}). For up and down
flavors, we have $n_u-n_{\bar u}=2/ V_{eff}$ and $n_d-n_{\bar
d}=1/ V_{eff}$
%\begin{equation}
%n_u-n_{\bar u}={2\over V_{eff}}, \,\,\,\, n_d-n_{\bar d}={1\over
%V_{eff}},
%\end{equation}
in  proton while $n_u-n_{\bar u}=1/ V_{eff}$ and $n_d-n_{\bar
d}=2/ V_{eff}$
%\begin{equation}
%n_u-n_{\bar u}={1\over V_{eff}}, \,\,\,\, n_d-n_{\bar d}={2\over
%V_{eff}},
%\end{equation}
in  neutron.

For the short-distance contribution, the two contributions  to the
self-energy of $s$-quark ($\bar s$) (a) and (b) are due to the
charged current ($W^{\pm}$) and neutral current  respectively, the
later is usually called as the tadpole-diagram\cite{Pal}.

The contribution due to the charged current is
\begin{equation}
\Sigma^{s}_1=\sqrt 2G_F\gamma^0L\sin^2\theta_C(n_u-n_{\bar u}),
\end{equation}
where $G_F$ is the Fermi coupling constant, $\theta_C$ is the
Cabibbo angle. The contribution due to the weak neutral current is
\begin{equation}
\Sigma^{s}_2=3\sqrt 2G_F(-1+{4\over
3}Q^{(s)}\sin^2\theta_{_W})\cdot
\sum_f(T_3^{(f)}-2Q^{(f)}\sin^2\theta_{_W})(n_f-n_{\bar f}),
\end{equation}
where $Q^{(f)}$ refers to the charge of corresponding quark ($u$,
$d$, $s$). Pal and Pham pointed that the axial part of the neutral
current does not contribute\cite{Pal}.

For the long-distance effects, in the calculations, we need an
effective vertex for $\bar sqM$ where $q$ can be either $u$ or
$d$-quarks and $M$ is a pseudoscalar or vector meson. Here we only
retain the lowest lying meson states such as $\pi,K,\rho$ etc. The
effective chiral Lagrangian for the interaction between quarks and
mesons has been derived by many authors\cite{Georgi,Yan}.

In terms of these effective vertices, the long-distance medium
correction to the  mass of strange quark can be evaluated and we
obtain
\begin{eqnarray} \label{m}
\Sigma_3^s &=& \gamma_0\frac{f^2_{kqs}}{2} [ (n_q-{n_{\bar q}}) +
\int \frac{d^3\mathbf{k}}{{(2\pi)}^3}
\frac{M_K^2}{m_s^2-2m_s \omega_k-M_K^2} f_F(\omega_k) \nonumber
\\& &-\int \frac{d^3\mathbf{k}}{{(2\pi)}^3} \frac{m_k^2}{m_s^2+2m_s
\omega_k-M_K^2}f_{F}(-\omega_k) ].
\end{eqnarray}
In order to avoid the pole in the second term  of Eq.~(\ref{m}),
we use the familiar Breit-Wigner formulation.

%\\
%\noindent (3)

%(i)
Our numerical results show that for the short-distance effects,
$\delta m= 92\ {\rm eV} \to 0.8\ {\rm KeV}$ for proton and $\delta
m=0.38\ {\rm KeV} \to 3.0\ {\rm KeV}$ for neutron,
%$$\delta m= 92\ {\rm eV} \sim 0.8\ {\rm KeV}, \;\;\;\; {\rm for\; proton},$$
%$$\delta m=0.38\ {\rm KeV} \sim 3.0\ {\rm KeV}, \;\;\;\; {\rm for\; neutron},$$
in the range of the effective nucleon radius $R\approx0.5 \to 1.0$
fm.

%(ii)
According to the picture of chiral field
theory\cite{eich,cheng,szcz}, the effective pseudovector coupling
implies $f_{kqs}=\frac{g_A}{\sqrt 2 f}$, where the axial-vector
coupling $g_A=0.75$. The pion decay constant $f_{\pi}=93$ MeV,
kaon decay constant $f_{K}=130$ MeV, for our estimation, an
approximate SU(3) symmetry might be valid, so that $f$ can be
taken as an average of $f_{\pi}$ and $f_K$. Thus we obtain $\delta
m\sim 10 \to 100$ MeV. One can trust that the order of the
effective coupling at the vertices  does not deviate too much from
this value. More detailed analysis can be found in
Ref.~10.
%\cite{Li:2001nv}.
%\\

%\noindent (4)
As a summary, we find that an asymmetry of the light quarks
%(u and d quarks) which exists
in nucleons can induce the expected asymmetry of the strange sea.
The short-distance effects are caused by the fundamental weak
interactions of the Standard Model, so that the corresponding
theoretical estimation of the asymmetry is more reliable, but due
to the heavy W(Z) bosons in the propagators, such effects can only
result in $\delta m$ of a few KeV. The main contribution to
$\delta m$ must come from the long-distance strong interaction, if
the phenomenological value of $\delta m$ is about 40 MeV as
determined by data. How to correctly evaluate such effects is the
key point, even though one can be convinced that the long-distance
effects should make a substantial contribution to $\delta m$.

In the history, there has been a dispute whether the parton
picture and the quark-meson interaction compromise with each
other, and if they do coincide, how to properly apply the picture
to evaluate phenomenological quantities is still an open %unsolved
problem. In this work, we just calculate the asymmetry of the
strange sea by this picture and obtain an estimate which meets the
value range from data fitting. Therefore we may consider that this
scenario has certain plausibility and its applicability should be
further tested in other calculations. The studies along this line
are worth more attention, because it is of obvious
significance for theory and phenomenological applications.
%\\

%\section*{Acknowledgements}

%This work is partially supported by the National Natural Science
%Foundation of China.

\end{document}